# Generation of Schubert polynomial series by nanophotonics

Hirotsugu Suzui[1], Kazuharu Uchiyama[1,*], Ryo Nakagomi[1], Hayato Saigo[2], Kingo Uchida[3], Makoto Naruse[4], and Hirokazu Hori[1]

[1] *University of Yamanashi, 4-3-11 Takeda, Kofu, Yamanashi 400-8511, Japan*

[2] *Nagahama Institute of Bio-Science and Technology, 1266 Tamura, Nagahama, Shiga 526-0829, Japan*

[3] *Ryukoku University, 1-5 Yokotani, Oe-cho, Seta, Otsu, Shiga 520-2194, Japan*

[4] *Department of Information Physics and Computing, Graduate School of Information Science and Technology, The Tokyo University, 7-3-1 Bunkyo-ku, Tokyo 113-8656, Japan*

[*] Corresponding author: kuchiyama@yamanashi.ac.jp

**Abstract**

Generation of irregular time series based on physical processes is indispensable in computing and artificial intelligence. In this report, we propose and experimentally demonstrate the generation of Schubert polynomials, which is the foundation of versatile permutations in mathematics, via optical near-field processes introduced in a photochromic crystal of diarylethene, which optical near-field excitation on the surface of a photochromic single crystal yields a chain of local photoisomerization, forming a complex pattern on the opposite side of the crystal. The incoming photon travels through the nanostructured photochromic crystal, and the exit position of the photon exhibits a versatile pattern. We experimentally generated Schubert matrices, corresponding to Schubert polynomials, via optical near-field density mapping. The versatility and correlations of the generated patterns could be reconfigured in either a soft or hard manner by adjusting the photon detection sensitivity. This is the first study of Schubert polynomial generation by physical processes or nanophotonics, paving the way toward future nano-scale intelligence devices and systems.

**Introduction**

Irregular time series play critical roles in information and communication technology today, including in secure information transfer [1, 2], Monte Carlo simulations [3], and machine learning [4, 5]. Physical processes in nature are interesting resources in providing irregular time series, including deterministic dynamics such as chaos [6] not just truly random sequences such by single photons [7]. Indeed, chaotic lasers enable interesting functionalities ranging from ultrafast random number generation [8] and photonic reservoir computing [9] to decision making and reinforcement learning [10]. Chaotic time series are remarkable in that their inherent correlations are determined by their initial conditions. A minute initial difference results in significantly different future trajectories, although the trajectories share common attributes specified by the dynamics therein. It has been found that the high complexity and high correlation of deterministic chaos contribute significantly to value alignment or decision making [10]. In general, it is important that a time series is complex and specified in certain conditions.

In this study, we demonstrated the physical generation of irregular time series from near-field optical systems using photoisomerization in a photochromic crystal, which were observed using a double-probe scanning near-field optical microscope (SNOM). Specifically, we generated Schubert matrices corresponding to Schubert polynomials [11, 12] by using a sequence of photon transmissions through a photochromic crystal photoisomerized in nanometer-scale. The spatial positions of the transmitted photons are related to the cell of Schubert polynomials. To our knowledge, this is the first report of the utilization of Schubert polynomials for physical random sequence generation.

**Optical near-field processes for complex systems**

The objective of this study was to produce highly correlated and yet significantly versatile time sequences, like those occurring in deterministic chaos dynamics. We have employed optical near-field processes on the nanometre scale, in which the physical system should obey certain restriction, corresponding to the inherent nonlinearity and sensitivity to initial conditions in the case of chaotic dynamics. Our previous studies [13, 14] revealed that local optical near-field excitation on the surface of a photochromic crystal yields local photoisomerization at a scale of 10 nm. In this case, photochromic reactions between colourless (transparent) open-ring isomer (**1o**) and blue-coloured closed-ring isomer (**1c**) proceeds in crystalline state, the coloured crystal upon UV irradiation was used as the recording material. The coloured material was decoloured due to cycloreversion reaction from **1c** to **1o** by optical near-field (Fig. 1a). Since the local transparency guides optical near field into the deeper layer, the local photoisomerization to the transparent state induces succeeding chain of local transparency including bifurcations, which leads to the formation of a complex pattern on the *opposite* surface of the crystal, as shown in Fig. 1b. Figure 1b indicates one of the experimentally observed pattern measured by double-probe optical near-field microscopy [13], in which near-field excitation was performed at a specific single point on one surface of the photochromic crystal by a probe while the probe being scanned on the opposite side to observe near-field optical transmission through the entire thickness of the crystal.

**Experiment**

Here we provide a brief review about the experiment reported in our previous study [13] to clarify the nature of the near field optical pattern shown in Fig. 1c. The sample used in the experiment was a photochromic single crystal of diarylethene (**1o**) with the molecular structure shown in Fig. 1a [15]. Upon UV irradiation, the sample converted to the blue coloured isomers, exhibiting visible light absorption with the peak wavelength in the green region. With visible light irradiation, the sample converted to the colourless isomers transparent to visible light. See *Methods* for details about the sample.

The measurements were conducted using a double-probe SNOM (UNISOKU, USM-1300S). The local optical excitation to the surface of diarylethene was performed with a metallic probe tip and optical near-field transmission measurement was conducted with an optical fibre probe (See *Methods* for details about the probes). The two probe tips were aligned so that their horizontal positions were separated less than 2 μm prior to the measurement [13]. The sample coloured by UV light was inserted between the two position-adjusted probes, and the approach of the two probes toward the sample was controlled via the scanning tunnelling microscopy (STM) method. The photoisomerization patterns generated by the local near-field light source at the metal probe tip were measured in a two-dimensional area of 2 μm × 2 μm with a resolution of 256 pixels × 256 pixels.

Figure 1c shows the resulting optical near-field intensity distribution after adopting a two-dimensional Gaussian filter with a standard deviation of 6 pixels, which corresponds to the resolution limit of the system. A complex structural pattern is observable, with a representative scale of 100–200 nm. Specifically, the local near-field excitation (by the metallic tip) does not uniformly propagate through the photochromic crystal, but a chain of local photoisomerization with a physical scale less than 1/5 of the optical wavelength was induced, involving spontaneous symmetry breaking, as mentioned in *Introduction*.

Such complex pattern generation has been considered to be due to a balance between the mechanical deformation of photochromic materials and photoisomerization [13, 14]; that is, local photoisomerization induces anisotropic deformation of molecular size, leading to mechanical anisotropic strain [16, 17], which induces subsequent photoisomerization in a non-uniform manner in the surrounding material (Fig. 2a). Hence, spontaneous symmetry breaking is evident in the local photoisomerization involving branching and selection, as schematically shown in Fig. 2b, leading to versatile pattern generation, as observed in Fig. 1c. In other words, complex *paths*, or transparent regions, are generated within the photochromic crystal.

Here we consider that an input light is irradiated via a near-field tip in a single photon level, whose photon energy is below that necessary to induce further photoisomerization. The photon travels through the generated complex transparent paths and is transmitted from a certain exit position on the opposite side of the input near-field probe, as schematically depicted in Fig. 2c. The spatial position of the output differs photon by photon. By regarding the spatial position as a different code, a sequence of irregular codes can be generated.

While the spatial position of the output photon is versatile, it is *not* arbitrary, because the transparent paths are produced in a specific manner in the path formation phase. Specifically, the geometry of the photon path in the photochromic crystal defines the underlying structure of the resulting code sequence. Such geometry in nano-optical structures corresponds to nonlinear dynamics in the case of chaotic sequence generation.

In this study, instead of conducting single photon measurements with an array of photodetectors, we experimentally observed the photon statistics at each of the spatial positions by using the SNOM. The experimental data provide probability distributions that emulate single photon arrival. Depending on the size of the single photon source needed for application, we rescaled the optical near-field intensity image. Figures 2d, 2e, and 2f show renormalized images with dimensions of 4 × 4, 8 × 8, and 16 × 16, respectively. Each pixel is a 125 nm square in the case of a 16 × 16 resolutions.

## Generation of Schubert polynomial series

Using this emulated photon source, we created what we call *Schubert matrices*. A Schubert matrix is a matrix that represents the one-to-one associations between row numbers (1 to $N$) and column numbers (1 to $N$) in an $N \times N$ matrix. When a cell ($i$, $j$) is 1, row $i$ and column $j$ are associated, and the other cells in the same row and column are 0. The total number of different Schubert matrices is $N!$. By replacing rows (or columns) of the matrix, the entire pattern is generated. In other words, a Schubert matrix is a permutation matrix with respect to a diagonal matrix. A Schubert matrix has a one-to-one correspondence with a Schubert polynomial through divided difference operators [11, 12]. See *Methods* for details about the definition of Schubert polynomials.

Figure 3 schematically shows how to generate a 4 × 4 Schubert matrix. Experimentally, we examined a square area with side lengths of 2 μm. The size of each pixel is about 500 nm in this configuration. The photon arrival probability is represented as an intensity distribution in Fig. 3a. (See *Methods* for details regarding the conversion from intensity into probability.) By using uniformly distributed pseudorandom numbers, the position of the first photon arrival was determined (Fig. 3b). Note that the maximum intensity position is not necessarily determined by the probabilistic nature of a single photon. According to the definition of a Schubert matrix, the probability of photon detection is configured as zero for pixels whose row or column is the same as that of the first photon arrival location (Fig. 3c). Based on the reconfigured probability distribution, the position of the second photon arrival was determined (Fig. 3d). Again, the photon detection probabilities of the pixels located in the same row or column as the second photon were set to zero (Fig. 3e), followed by the determination of the third photon arrival position (Fig. 3f). The position of the fourth photon was determined automatically (Fig. 3g), because no two rows or columns could have photons. The resulting Schubert matrix is presented in Fig. 3h. This pattern is depicted by [2134], showing the column numbers of the photon detection pixels (coloured in yellow) in the order of the rows. The corresponding Schubert polynomial is $x_1$. (See *Methods* for details regarding Schubert polynomials.) There are 4! = 24 kinds of Schubert polynomials in total in the case of a 4 × 4 Schubert matrix, which is not a large number. However, the total number of possible polynomials grows exponentially as the size of the Schubert matrix increases. For a 16 × 16 Schubert matrix, there are 16! kinds of polynomials, which is on the order of $10^{13}$. Thus, extremely versatile sequences are physically generated by optical near-field processes.

We generated Schubert matrix series with dimensions of 16 × 16 using the above-mentioned procedures. The side length of each cell is about 125 nm. According to our previous research [13, 14], the size of elemental photoisomerization is less than 100 nm; hence, a single pixel corresponds to a few terminals of the light paths at most. Figure 4 summarizes the results in the case of 16 × 16 resolution. Based on the SNOM image shown in Fig. 4a, 10,000 Schubert matrices were generated. Figure 4b depicts the first 10 matrices, where versatile patterns are observable. Figure 4c represents the mean of these 10,000 matrices. It should be noted that the mean matrix and the original SNOM image have structural similarity, but the intensity distributions are very different because of the exclusive properties, meaning that only a single pixel occupies a row and column in the process of Schubert matrix generation. Therefore, for example, if an SNOM image contains multiple high intensity pixels in a common column or row, the resulting Schubert matrix exhibits a versatile pattern.

As discussed above, the arrival of a single photon immediately provides an element in the Schubert matrix. By implementing an additional constraint on the photon detection, a Schubert matrix that is strictly correlated to the source SNOM pattern can be generated. We set a threshold for the photon detection; that is, if the number of photons exceeded a certain threshold, an element of the Schubert matrix was determined. In this case, the high intensity positions in SNOM images are highly likely to yield elements in the Schubert matrix. That is, the properties in the photochromic nanostructure are more strongly transferred to the resulting Schubert matrix.

Figures 4d and 4e present the mean Schubert matrices from 10,000 matrices when the threshold level was set to 3 and 5 photons, respectively. Whereas the maximum value in the mean Schubert matrix is about 0.1 when the threshold is 1, the corresponding value increases to 0.2 and 0.4 in the cases of 3 and 5 photons, respectively. That is to say, a specific element is selected more often in the Schubert matrix as the threshold photon number increases. Therefore, the correlation between the optical near-field distributions and the resulting sequences can be tuned by the degree of photon detection; highly sensitive detection provides higher randomness or *softness*, while lower sensitivity yields limited versatility or *hardness*. It should be noted that the same optical near-field processes and the same measurement apparatus provide tunable or reconfigurable functionality, which would be useful for applications such as large-scale decision making [18].

We examined the versatility of the Schubert matrix, or its corresponding time sequence, when the SNOM image was rescaled as 8 × 8 because the maximum number of patterns is reasonable in that case (8! = 40,320). The number of unique Schubert matrices decreases as the photon number threshold increases. Figure 5a shows the number of unique Schubert matrices as a function of repetition cycle where, after 10,000 repetitions, the number of unique patterns is only 4925 and 1809 when the threshold is 5 and 10 photons, respectively. In the case of the Schubert matrices created by pseudorandom numbers, there were 8768 unique patterns out of 10,000 matrices.

Figure 5b summarizes mean correlations between pairs among all 10,000 Schubert matrices. The correlation is nearly zero when the photon threshold is 1, while the correlation increases as the photon threshold increases; the correlation is more than 0.15 and 0.3 when the threshold is 10 and 20 photons, respectively. That is, tunable correlation was successfully accomplished by controlling the sensitivity of photon detection. It should be emphasized that, although the resulting patterns are extremely diverse, especially in the case of lower photon sensitivity, they are highly regulated by the SNOM images. In other words, they are not purely random, as is clearly evident from the mean structure of the resulting Schubert matrix in Fig. 4c.

**Conclusion**

We proposed and experimentally demonstrated the generation of Schubert polynomials via optical near-field processes. The local photoisomerization induced in a photochromic crystal provides a complex, non-uniform internal structure in the crystal, leading to a variety of photon transmission pathways. We employed a two-dimensional photon observation strategy; by introducing non-sensitive portions into the measurements concerning the definition of Schubert polynomials, versatile patterns were successfully generated. We also demonstrated that, by changing the photon detection sensitivity, the variety of the resulting sequences as well as their correlations with one another can be configured. In other words, the relations between the generated Schubert polynomials and the nanostructure properties can be tuned, which is important in future applications in which autonomous adaptation to the environment is crucial, such as in decision making and soft robotics [18, 19].

The light transmission through the complex paths in a nanomaterial involves a quantum structure; a photon travels the superposition of multiple paths but converges to a certain single observation. Thus, this system would be an interesting and important platform for quantum measurement [20]. Furthermore, *multiple* local input excitations of the material, unlike the *single* local excitation approach employed in the present study, would provide higher functionality, such as enhanced versatility. As discussed in *Introduction*, the demand for irregular time series is increasing for use in information security and artificial intelligence, among other fields. The fusion of advanced mathematics and physical processes including nanophotonics would be an interesting and exciting interdisciplinary topic for future research.

**Methods**

**Sample preparation.** The single crystal used in this study had a square cross-section with a side length of approximately 0.5 mm and 0.1 mm thickness. On both sides of the sample, a thin Pt layer approximately 10 nm thick was coated to make the surface conductive, because the position of our optical near-field probe tip was controlled by utilizing a STM.

**Probe preparation.** A metallic probe tip for local optical excitation was sharpened using an electropolishing method to a radius of curvature of several tens of nanometres and coated with an approximately 20-nm-thick Au layer. This metal probe was illuminated by laser light of wavelength 532 nm with an intensity of 1 μW/cm$^2$ to generate optical near-field by local electric field enhancement at the tip. The optical fibre probe was obtained by sharpening the optical fibre using a buffered hydrofluoric acid solution followed by 10-nm-thick Pt coating.

**Schubert polynomial [11, 12].** The four-dimensional Schubert matrix (that is, the four-dimensional symmetric group $S_4$) and the corresponding Schubert polynomial ($\mathfrak{S}_\omega, \omega \in S_4$) are explained below as an example. A four-dimensional identity matrix can be expressed as [1234] (*id*). The anti-identity matrix [4321] ($\omega_0$) is the permutation of the longest length (6) in $S_4$ ("length" is the number of necessary cycles of simple transpositions to generate the permutation from *id*). The Schubert polynomial for $\omega_0$ is $\mathfrak{S}_{\omega_0} \equiv x_1^3 x_2^2 x_3$ (in the *n*-dimensional Schubert matrix, $\mathfrak{S}_{\omega_0} \equiv x_1^{n-1} x_2^{n-2} \cdots x_{n-2}^2 x_{n-1}$). Other polynomials corresponding to matrices $\omega$ are determined by divided difference operators corresponding to necessary simple transpositions to convert to the anti-diagonal matrix ($\omega^{-1}\omega_0$). One matrix [4312] becomes [4321] by exchanging rows 3 and 4 (this simple transposition is represented as $s_3$). Thus, the Schubert polynomial corresponding to [4312] is $\partial_{\omega^{-1}\omega_0}\mathfrak{S}_{\omega_0} = \partial_3 \mathfrak{S}_{\omega_0} = (\mathfrak{S}_{\omega_0} - s_3 \mathfrak{S}_{\omega_0})/(x_3 - x_4) = (x_1^3 x_2^2 x_3 - x_1^3 x_2^2 x_4)/(x_3 - x_4) = x_1^3 x_2^2$. Other examples include $\mathfrak{S}_{id} = 1$, $\mathfrak{S}_{1432} = \partial_1 \partial_2 \partial_3 \mathfrak{S}_{\omega_0} = x_1^2 x_2 + x_1^2 x_3 + x_1 x_2^2 + x_1 x_2 x_3 + x_2^2 x_3$, and $\mathfrak{S}_{2431} = \partial_1 \partial_2 \mathfrak{S}_{\omega_0} = x_1^2 x_2 x_3 + x_1 x_2^2 x_3$.

**Conversion of optical near-field intensity map into probability map.** Let the intensity distribution of the obtained near-field light be *NF* (*x*, *y*) (*x* = 1, ..., *N*, *y* = 1, ..., *N*, where *N* is the pixel size of the SNOM image). The value obtained by subtracting 0.99 times the minimum intensity is defined as *NFS* (*x*, *y*) = *NF* (*x*, *y*) − min (*NF* (*x*, *y*)) × 0.99. Since the near-field light intensity contains background noise, 0.99 times the minimum value is subtracted as noise in this case. The probability distribution is given by *P* (*x*, *y*) = *NFS* (*x*, *y*) / Σ *NFS* (*x*, *y*). In addition, for pixels whose sensitivity is made 0 in the process of deriving the Schubert matrix, perform the update *NFS* (*x*, *y*) = 0 and recalculate *P* (*x*, *y*) = *NFS* (*x*, *y*) / Σ *NFS* (*x*, *y*).

**Acknowledgement**

This work was supported in part by the CREST program (JPMJCR17N2) of the Japan Science and Technology Agency and the Grants-in-Aid for Scientific Research (JP26107012, JP17H01277, JP25286067) from the Japan Society for the Promotion of Science. The authors thank Izumi Ojima and Kazuya Okamura for the valuable discussions.

**Author Contributions**

H.H., K.U.3 and M.N. directed the project. H.S., K.U.1, R.N. and H.H. designed the experimental strategy and methods. K.U.3 prepared the photochromic materials. H.S., K.U.1, R.N. and H.H. performed near-field optics experiments. H.S., K.U.1, M.N., H.S. and H.H. analysed the data. H.S., K.U.1 and M.N. wrote the paper, and all of the authors contributed to the preparation of the manuscript.

**Competing Interests**

The authors declare no competing interests.

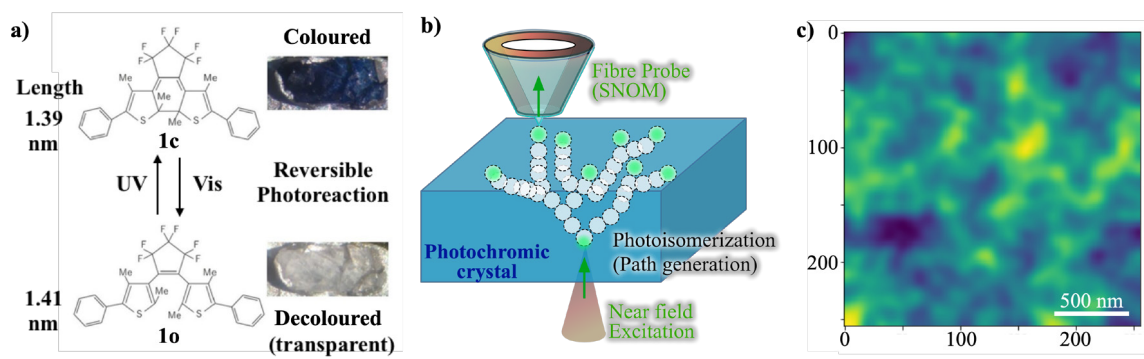

**Fig. 1 Double-probe SNOM measurement of photons propagating through the transparent paths in a photochromic crystal. a:** Diarylethene molecule structure and photoisomerization used in the present study. **b:** Experimental setup and schematic image of photon detection by double probe. **c:** Map of optical near-field intensity.

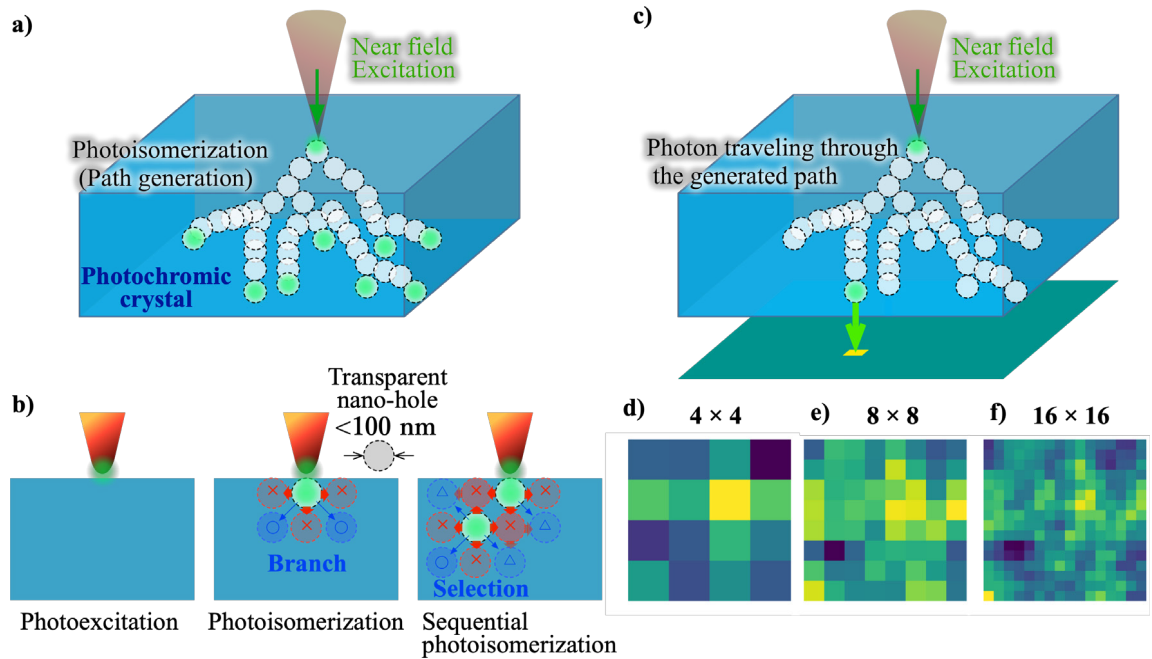

**Fig. 2 Photon source for generation of Schubert calculus matrices with nanometre-scale photoisomerization in photochromic crystal and double-probe SPM measurements. a:** Formation of transparent path by photoisomerization with spontaneous symmetry breaking. **b:** Chain of anisotropic local photoisomerizations from the local photoexcitation. **c:** Detection of photons propagating along the transparent path in a photochromic crystal for the generation of a Schubert matrix. **d–f:** 4 × 4, 8 × 8, and 16 × 16 matrices generated by coarse graining of optical near-field images as photon detection probability density maps.

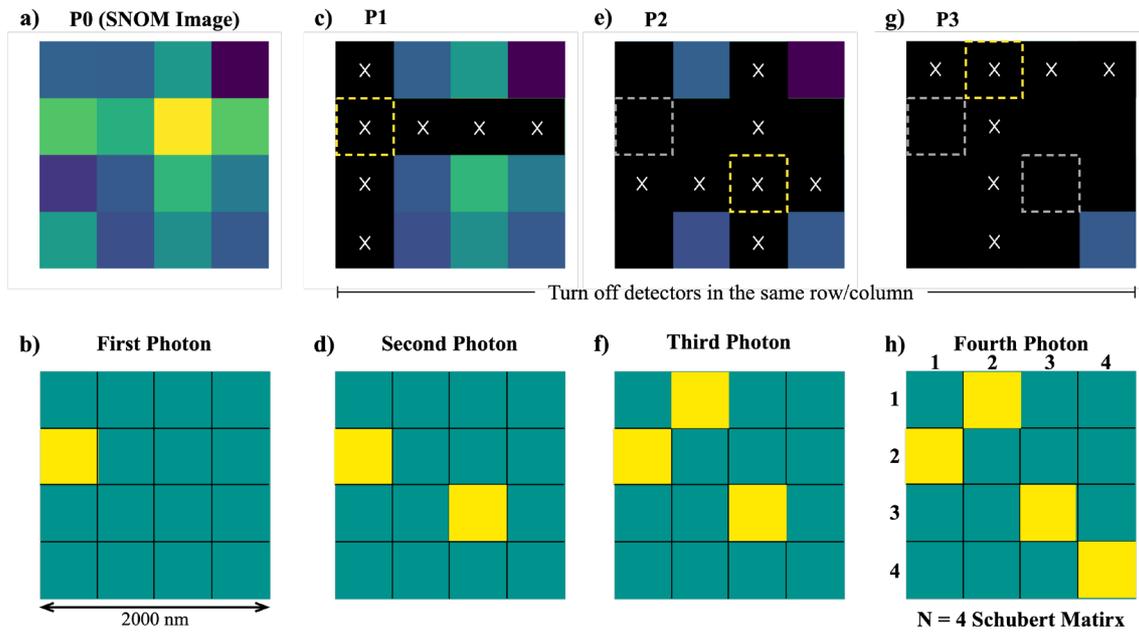

**Fig. 3 Method of Schubert matrix generation using an optical near-field intensity image as a probability density map of photon detection. a:** 4 × 4 optical near-field intensity image used as a probability map. **b:** First photon detected, whose position was stochastically determined using the probability map (represented by a yellow pixel). **c**: Setting of the probability of points in the same row/column as the first photon to zero (black). **d–g:** Process of sequential photon detection and probability changes. **h:** Schubert matrix generated through processes **a**–**g**.

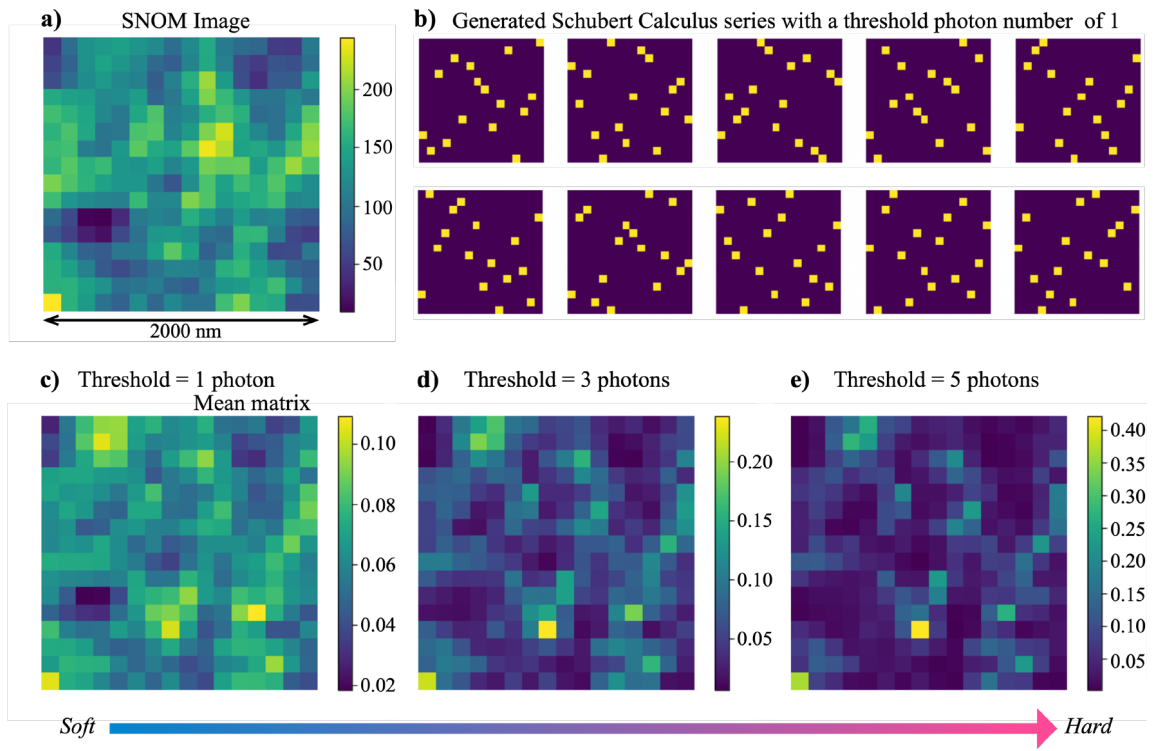

**Fig. 4 Generation of Schubert matrix series with different threshold photon numbers. a:** Observed SNOM image (the same as in Fig. 2e). **b:** First 10 generated Schubert matrices with a threshold photon number of 1. **c–e:** Mean matrix of the 10,000 Schubert matrices generated with different threshold photon numbers: 1 photon for **c**, 3 photons for **d**, and 5 photons for **e**.

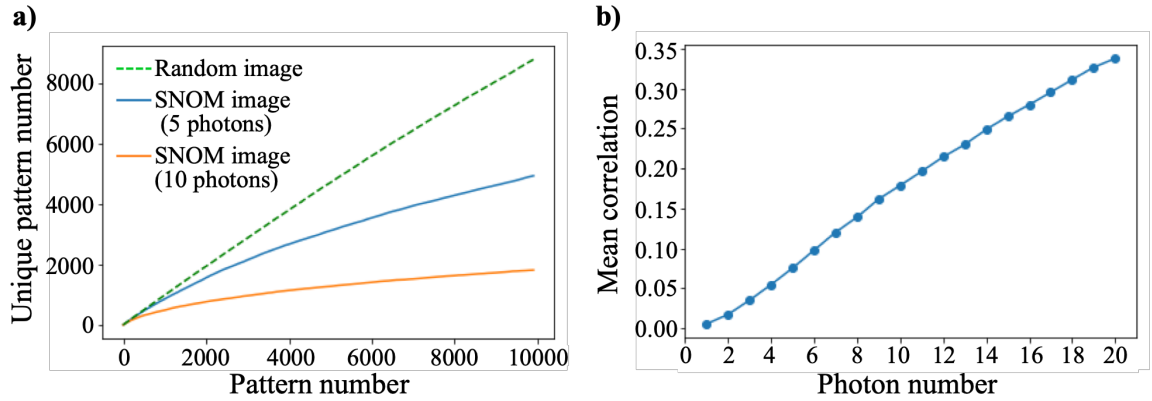

**Fig. 5 Properties of the generated Schubert matrices**. **a:** Time evolution of the number of unique Schubert matrices as a function of repetition cycle. The results obtained with thresholds of 5 photons and 10 photons are represented by blue and red lines, respectively. The corresponding results for randomly generated matrices are also presented, as a dotted line. **b:** Mean correlations between pairs of all 10,000 Schubert matrices for different thresholds, from 1 photon to 20 photons.